\documentstyle[preprint,amsmath,aps,epsf]{revtex}

\begin{document}
%\draft 

\title{Innermost circular orbit of binary black holes\\ at the third
post-Newtonian approximation} \author{Luc Blanchet}
\address{Gravitation et Cosmologie (GReCO),\\ Institut d'Astrophysique
de Paris -- C.N.R.S.,\\ 98\textsuperscript{~\!bis} boulevard Arago,
75014 Paris, France} \date{\today}
%\twocolumn[
\maketitle
\widetext
\begin{abstract}
The equations of motion of two point masses have recently been derived
at the 3PN approximation of general relativity. From that work we
determine the location of the innermost circular orbit or ICO, defined
by the minimum of the binary's 3PN energy as a function of the orbital
frequency for circular orbits. We find that the post-Newtonian series
converges well for equal masses. Spin effects appropriate to
corotational black-hole binaries are included. We compare the result
with a recent numerical calculation of the ICO in the case of two
black holes moving on exactly circular orbits (helical symmetry). The
agreement is remarkably good, indicating that the 3PN approximation is
adequate to accurately locate the ICO of two black holes with
comparable masses. This conclusion is reached with the post-Newtonian
expansion expressed in the standard Taylor form, without using
resummation techniques such as Pad\'e approximants and/or
effective-one-body methods.
\end{abstract}

\pacs{04.30.-w, 04.80.Nn, 97.60.Jd, 97.60.Lf}
%]

%\narrowtext

The aim of this article is to compute the innermost circular orbit
(ICO) of point-particle binaries in post-Newtonian approximations, and
to compare the result with numerical simulations. For the present
purpose, the ICO is defined as the minimum, when it exists, of the
binary's energy function $E(\Omega)$ of circular orbits, where
$\Omega$ denotes the orbital frequency. The definition is motivated by
our comparison with the numerical work, because this is precisely that
minimum which is computed numerically\footnote{In particular, we do
not define the ICO as a point of dynamical (general-relativistic)
unstability.}. The energy function is given by the invariant quantity
associated with the {\it conservative} part of the post-Newtonian
dynamics, i.e. ignoring the radiation reaction effects. It can be
argued, because the radiation reaction damping is neglected, that the
importance of the ICO does not lie so much on its strong physical
significance, but on the fact that it represents a very useful
reference point on the definition of which the post-Newtonian and
numerical methods agree.

The question of the conservative dynamics of compact binary systems
has been resolved in recent years at the third post-Newtonian (3PN)
approximation, corresponding to the order $1/c^6$ beyond the Newtonian
force. After the previous work of Refs. \cite{OO,S85}, Jaranowski and
Sch\"afer \cite{JaraS}, and Damour, Jaranowski and Sch\"afer
\cite{DJS}, have applied at 3PN order the ADM-Hamiltonian formalism of
general relativity. On the other hand, extending the method of
Ref. \cite{BFP98}, Blanchet and Faye \cite{BF,BFreg} performed a 3PN
iteration of the equations of motion (instead of a Hamiltonian) in
harmonic coordinates. In the latter approaches, the compact objects
are modelled by point particles, described solely by two mass
parameters $m_1$ and $m_2$. The end results are physically equivalent
in the sense that there exists a unique transformation of the
particle's dynamical variables that changes the 3PN
harmonic-coordinates Lagrangian of de Andrade, Blanchet and Faye
\cite{ABF01} into another Lagrangian, whose Legendre transform is
identical with the 3PN ADM-coordinates Hamiltonian of Damour,
Jaranowski and Sch\"afer \cite{DJS}.

Post-Newtonian computations of the motion of point particles face the
problem of the regularization of the infinite self-field of the
particles. The regularization scheme of Hadamard, in ``standard''
form, was originally adopted in the ADM-Hamiltonian approach
\cite{JaraS}. Then an ``improved'' version of this regularization was
defined in Refs. \cite{BFreg} and applied to the computation of the
harmonic-coordinates equations of motion \cite{BF}.  Unfortunately, it
has been shown that the Hadamard regularization, either in standard or
improved form, leaves unspecified one and only one numerical
coefficient in the 3PN equations of motion, $\omega_{\rm static}$ in
the ADM-Hamiltonian approach, $\lambda$ in the harmonic-coordinates
formalism. The parameter $\omega_{\rm static}$ can be seen as due to
some ``ambiguity'' of the standard Hadamard regularization, while
$\lambda$ appears rather like a parameter of ``incompleteness'' in the
improved version \cite{BFreg} of this regularization. However, these
constants turned out to be equivalent, in the sense that
\cite{BF,DJS,ABF01}

\begin{equation}\label{1}
\lambda = -\frac{3}{11} \omega_{\rm static}-\frac{1987}{3080}\;.
\end{equation}
It has been argued in Ref. \cite{DJSisco} that the numerical value of
$\omega_{\rm static}$ could be $\simeq -9$, because for such a value
some different ``resummation'' techniques, when they are implemented
at the 3PN order, give approximately the same numerical result for the
ICO. Even more, it was suggested \cite{DJSisco} that $\omega_{\rm
static}$ might be precisely equal to $\omega_{\rm
static}^*=-\frac{47}{3}+\frac{41}{64}\pi^2\simeq -9.34$ (corresponding
to $\lambda^*\simeq 1.90$). But, more recently, a computation of
$\omega_{\rm static}$ has been performed by means of a dimensional
regularization, instead of the Hadamard regularization, within the
ADM-Hamiltonian formalism \cite{DJSdim}, with result

\begin{equation}\label{1'}
\omega_{\rm
static}=0~~\Longleftrightarrow~~\lambda=-\frac{1987}{3080}\simeq
-0.64\;.
\end{equation}
We adopt in this article the latter value as our preferred one, but in
fact it is convenient to keep the ambiguity parameter unspecified, and
to investigate the behaviour of the solutions for different values of
$\lambda$ or $\omega_{\rm static}$. For instance, we shall keep an eye
on the values $\omega_{\rm static}^*\simeq -9.34$ and also
$\lambda=0\Longleftrightarrow\omega_{\rm static}\simeq -2.37$. The
latter case corresponds to the special instance where certain
logarithmic constants associated with the Hadamard regularization in
harmonic coordinates do not depend on the masses \cite{BF}. Notice
that the result (\ref{1'}) is quite different from $\omega_{\rm
static}^*\simeq -9.34$~: this suggests, according to
Ref. \cite{DJSisco}, that different resummation techniques, {\it viz}
Pad\'e approximants \cite{DIS98} and effective-one-body methods
\cite{BD}, which are designed to ``accelerate'' the convergence of the
post-Newtonian series, do not in fact converge toward the same
``exact'' solution (or, at least, not as fast as expected).

Let us now compute the ICO of two point particles (modelling black
holes) at the 3PN order thanks to the previous body of works
\cite{OO,S85,JaraS,DJS,BFP98,BF,BFreg,ABF01}. The circular-orbit
binding energy $E$ (in the center-of-mass frame), and angular-momentum
$J$, are deduced either from the 3PN harmonic-coordinates Lagrangian
\cite{ABF01} or, equivalently, from the 3PN ADM-coordinates
Hamiltonian \cite{DJS} (we neglect the 2.5PN radiation damping). These
functions are expressed in invariant form (the same in different
coordinate systems), i.e. with the help of the angular orbital
frequency $\Omega$. The 3PN energy (per unit of total mass $M$),
describing ``irrotational'' circular-orbit binaries, is

\begin{eqnarray}\label{2}
\frac{E(\Omega)}{M} &=& -\frac{\nu}{2}\left(M \Omega\right)^{2/3}
 \biggl\{ 1 +\left(-\frac{3}{4}-\frac{\nu}{12}\right) \left(M
 \Omega\right)^{2/3}+ \left(-\frac{27}{8}+\frac{19}{8}\nu
 -\frac{\nu^2}{24}\right) \left(M \Omega\right)^{4/3} \nonumber\\
 &&+\left(-\frac{675}{64}+\left[\frac{209323}{4032}-\frac{205}{96}\pi^2
 -\frac{110}{9}\lambda\right]\nu-\frac{155}{96}\nu^2
 -\frac{35}{5184}\nu^3\right)\left(M \Omega\right)^{2} \biggr\}\;.
\end{eqnarray} 
All over this paper we pose $G=c=1$. Mass parameters are $M=m_1+m_2$,
and the symmetric mass ratio $\nu=m_1m_2/M^2$ such that $0<\nu\leq
\frac{1}{4}$, with $\nu=\frac{1}{4}$ in the equal-mass case and
$\nu\to 0$ in the test-mass limit for one of the bodies. The 3PN
angular momentum, scaled by $M^2$, reads

\begin{eqnarray}\label{3}
\frac{J(\Omega)}{M^2}&=& \nu \left(M \Omega\right)^{-1/3} \biggl\{ 1
 +\left(\frac{3}{2}+\frac{\nu}{6}\right) \left(M \Omega\right)^{2/3}
 + \left(\frac{27}{8}-\frac{19}{8}\nu +\frac{\nu^2}{24}\right)
 \left(M \Omega\right)^{4/3} \nonumber\\
 &&+\left(\frac{135}{16}+\left[-\frac{209323}{5040}+\frac{41}{24}\pi^2
 +\frac{88}{9}\lambda\right]\nu+\frac{31}{24}\nu^2
 +\frac{7}{1296}\nu^3\right)\left(M \Omega\right)^{2} \biggr\}\;.
\end{eqnarray}
The variations of the energy and angular momentum of the binary on the
circular orbit during the inspiral phase obey the evolutionary (or
``thermodynamic'') law

\begin{equation}\label{3'}
\frac{dE}{d\Omega}=\Omega \frac{dJ}{d\Omega}\;,
\end{equation}
which is equivalent, {\it via} the energy and angular-momentum balance
equations, to the same relation but between the corresponding
gravitational-wave fluxes at infinity. From Eq. (\ref{3'}), we see
that the points of extremum for $E$ and $J$ are the same. In the limit
$\nu\to 0$, Eqs. (\ref{2})-(\ref{3}) reduce to the 3PN approximations
of the known energy and angular momentum of a test particle in the
Schwarzschild background~:

\begin{mathletters}\label{3''}\begin{eqnarray}
\frac{E^{\rm Sch}(\Omega)}{M}&=&\nu\left\{\left(1-2(M\Omega)^{2/3}\right)
\left(1-3(M\Omega)^{2/3}\right)^{-1/2}-1\right\}\;,\label{3a''}\\
\frac{J^{\rm Sch}(\Omega)}{M^2}&=&\nu
(M\Omega)^{-1/3}\left(1-3(M\Omega)^{2/3}\right)^{-1/2}\;.
\end{eqnarray}\end{mathletters}
We recall that in this case the location of the ICO is given by $M
\Omega^{\rm Sch}_{\rm ICO}=6^{-3/2}$, with $E^{\rm Sch}_{\rm ICO}=\nu
M\left(\sqrt{\frac{8}{9}}-1\right)$ and $J^{\rm Sch}_{\rm ICO}=\nu M^2
\sqrt{12}$.

The straightforward post-Newtonian method we follow in this article
can be justified by the following arguments. At the location of the
ICO we shall find that $M \Omega_{\rm ICO}$ is of the order of
10\%. Therefore, we expect that the 1PN approximation will grossly
correspond to a relative modification of the binding energy of the
order of $v^2\sim (M \Omega_{\rm ICO})^{2/3}$ i.e. 20\%; and similarly
that the 2PN and 3PN approximations will yield some effects of
magnitude about 5\% and 1\% respectively. Consequently the
post-Newtonian method should be adequate in the regime of the ICO,
provided that it is implemented up to the 3PN order, so as to be
accurate enough. On the other hand, we see that the 1PN order should
yield a rather poor estimate of the position of the ICO.

Let us now confirm these estimates with the numerical values for the
post-Newtonian coefficients in the energy function (\ref{2}). As we
see from TABLE \ref{tab0}, in the case of comparable masses and of our
preferred value (\ref{1'}) for the ambiguity parameter, the absolute
values of the post-Newtonian coefficients are roughly of the order of
one (they do not apparently increase with the order of
approximation). This means that the previous estimates are essentially
correct. In particular the 3PN approximation should be close to the
``exact'' value for the ICO. The post-Newtonian series seems to
``converge well'' (in the case where $\nu=\frac{1}{4}$ and
$\omega_{\rm static}=0$), with a ``convergence radius'' of the order
of one, i.e. at a much higher frequency than the frequency of the
ICO\footnote{Actually the post-Newtonian series could be only
asymptotic (hence divergent), but nevertheless it should give good
results provided that the series is truncated near some optimal order
of approximation. In this article we assume that 3PN is not too far
from that optimum.}. By contrast, we recover in TABLE I the well-known
result (see e.g. \cite{3mn,P95}) that in the perturbative case $\nu\to
0$ the post-Newtonian series converges slowly~: the coefficients
increase roughly by a factor 3 at each post-Newtonian order,
reflecting the fact that the radius of convergence of the series is
$\frac{1}{3}$. This is clear from the exact expression (\ref{3a''}),
in which the pole at the value $\frac{1}{3}$ corresponds to the light
ring of the Schwarzschild metric. Thus the post-Newtonian method is
not very appropriate to the case $\nu=0$, where even the 3PN order
would rather poorly approximate the ICO. The situation is therefore
the following~: in the case of comparable masses, we do not have the
exact solution, but fortunately the straightforward post-Newtonian
approach is expected to be accurate; in the perturbative limit
$\nu=0$, the post-Newtonian series is poorly convergent, but gladly
this does not matter because we know the exact results (\ref{3''}).

\begin{figure}[htbp]
\centerline{\epsfxsize=14cm \epsfbox{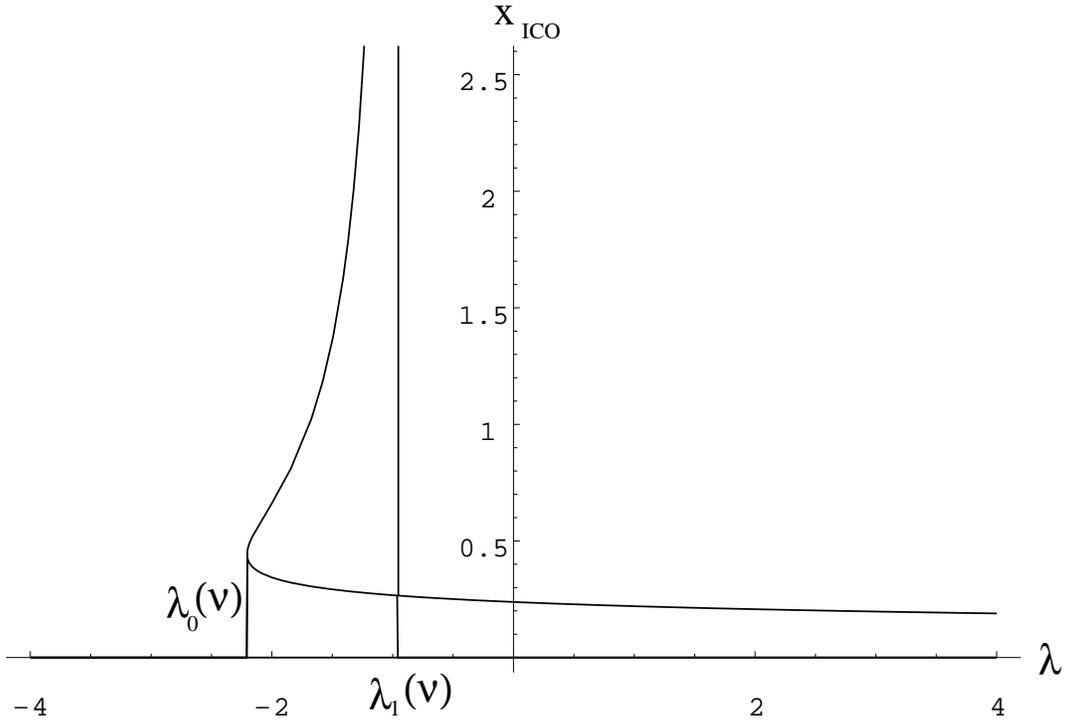}}
\vspace{0.5cm}
\caption{The possible solutions as a function of the regularization
constant $\lambda$. There is no solution when
$\lambda<\lambda_0(\nu)$, two possible solutions when
$\lambda_0(\nu)\leq\lambda<\lambda_1(\nu)$ [which become degenerate at
$\lambda=\lambda_0(\nu)$], and a unique solution when
$\lambda_1(\nu)\leq\lambda$. The upper branch, existing between
$\lambda_0(\nu)$ and the vertical asymptote at
$\lambda=\lambda_1(\nu)$, is actually a maximum of the energy.}
\label{fig1}
\end{figure}

Having thus justified the validity of our approximation, we look for
the point at which both $E(\Omega)$ and $J(\Omega)$ take some minimal
values $E_{\rm ICO}=E(\Omega_{\rm ICO})$ and $J_{\rm
ICO}=J(\Omega_{\rm ICO})$. As we see from Eq. (\ref{2}), at the 3PN
order $E(\Omega)$ is a polynomial of the fourth degree in the
frequency-parameter $x\equiv \left(M \Omega\right)^{2/3}$. Therefore,
the value of the minimum, $x_{\rm ICO}=\left(M \Omega_{\rm
ICO}\right)^{2/3}$, must be a real positive solution of an algebraic
equation of the third degree (in general)~:

\begin{equation}\label{4}
1+\alpha x+\beta
x^2+\gamma x^3 = 0\;.
\end{equation}
The coefficients are straightforwardly obtained from Eq. (\ref{2}) as

\begin{mathletters}\label{5}\begin{eqnarray}
\alpha(\nu)&=&-\frac{3}{2}-\frac{\nu}{6}\;,\label{5a}\\
\beta(\nu)&=&-\frac{81}{8}+\frac{57}{8}\nu
-\frac{\nu^2}{8}\;,\label{5b}\\
\gamma(\nu,\lambda)&=&-\frac{675}{16}
+\left[\frac{209323}{1008}-\frac{205}{24}\pi^2
-\frac{440}{9}\lambda\right]\nu-\frac{155}{24}\nu^2
-\frac{35}{1296}\nu^3\;.\label{5c}
\end{eqnarray}\end{mathletters}
The regularization constant $\lambda$ enters only the third-degree
monomial (3PN order). Let us describe, in a qualitative way, the
existence of solutions of Eq. (\ref{4}). We find that the equation
does not always admit a unique real positive solution, nor even
several of them. This depends, for a given choice of the mass ratio
$\nu$, on the constant $\lambda$. When $\lambda$ happens to be smaller
that some ``critical'' value $\lambda_0(\nu)$, depending on $\nu$,
there is {\it no} (real positive) solution, and therefore there is no
ICO at the 3PN order. When $\lambda$ is between $\lambda_0(\nu)$ and
another ``critical'' value $\lambda_1(\nu)$, also depending on $\nu$,
we obtain {\it two} real positive solutions. In this case, the energy
function admits two extrema, a minimum and a maximum. The maximum
occurs at a higher frequency than the minimum of the ICO, and is to be
discarded on physical grounds (the corresponding frequency is
generally too high, e.g. higher than $M^{-1}$, for being of physical
interest). Finally, when $\lambda$ is larger than $\lambda_1(\nu)$,
there is one and only one real positive solution~: $x_{\rm ICO}$, and
this is a minimum of the energy. The latter regime, where the
circular-orbit energy admits a unique extremum, which is a minimum
(like for the Schwarzschild metric), is the simplest on the physical
point of view. The interesting values of $\lambda$ are located in the
regime where $\lambda\geq \lambda_1(\nu)$ (for irrotational
binaries). We summarize our discussion in FIG. \ref{fig1}.

\begin{figure}[htbp]
\centerline{\epsfxsize=14cm \epsfbox{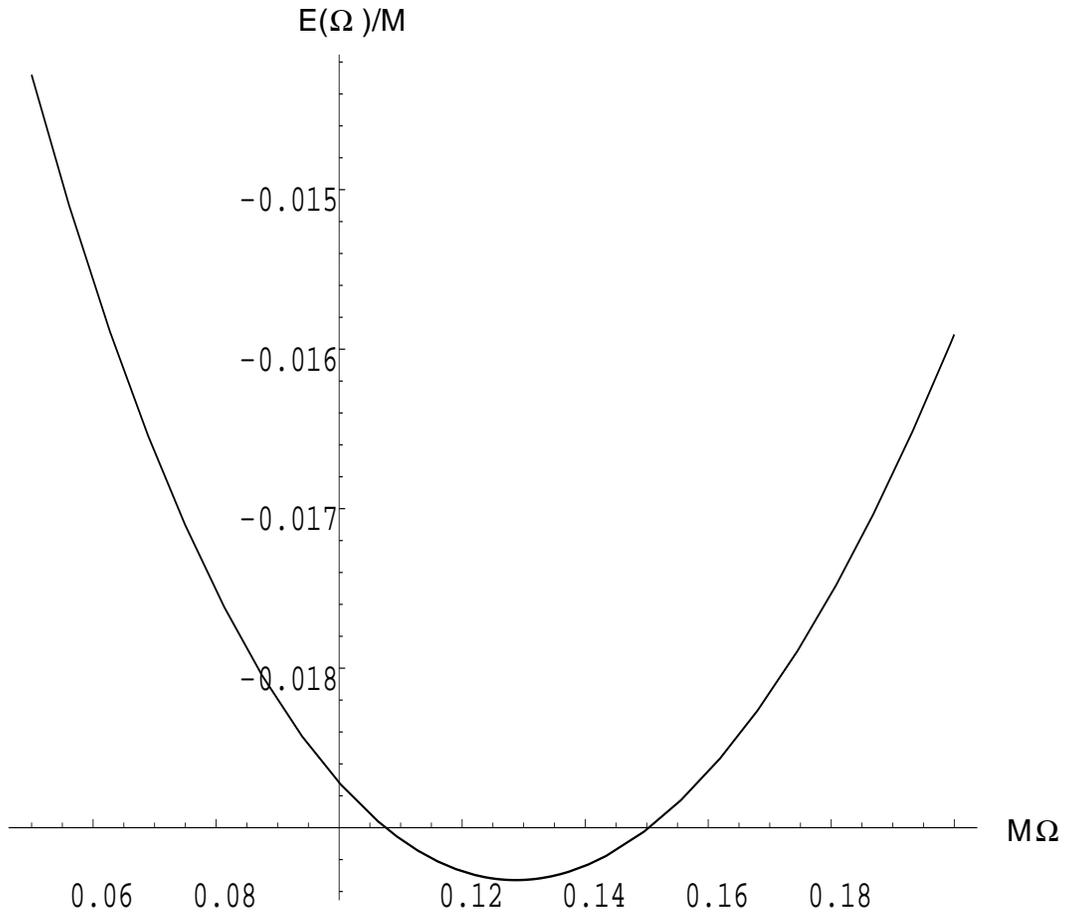}}
\vspace{0.5cm}
\caption{The 3PN energy function $E(\Omega)$ for equal-mass binaries
and $\omega_{\rm static}=0$.}
\label{fig2}
\end{figure}

It is not difficult to determine analytically the functions
$\lambda_0(\nu)$ and $\lambda_1(\nu)$. Indeed, $\lambda_0(\nu)$
represents simply the minimal value of the function $x_{\rm
ICO}\rightarrow\lambda(\nu,x_{\rm ICO})$ (see
FIG. \ref{fig1}). Using also Eq. (\ref{4}), we readily find the
mathematical relation defining $\lambda_0(\nu)$~:

\begin{equation}\label{6}
\lambda=\lambda_0(\nu)~~\Longleftrightarrow~~\gamma(\nu,\lambda)
=\frac{2}{27} \left[[\alpha^2(\nu)-3\beta(\nu)]^{3/2}-\alpha^3(\nu)
+\frac{9}{2}\alpha(\nu)\beta(\nu)\right]\;,
\end{equation}
from which the explicit expression of $\lambda_0(\nu)$ can be found
using Eqs. (\ref{5}). On the other hand, the function $\lambda_1(\nu)$
is determined by the cancellation of the third-degree coefficient in
the equation (\ref{4}), i.e.

\begin{equation}\label{7}
\lambda=\lambda_1(\nu)~~\Longleftrightarrow~~\gamma(\nu,\lambda)=0\;.
\end{equation}
The expression of $\lambda_1(\nu)$ then follows from using
Eq. (\ref{5c}). For allowed values of $\nu\in
\big]0,\frac{1}{4}\big]$, we find that both $\lambda_0(\nu)$ and
$\lambda_1(\nu)$ are increasing functions of $\nu$, with maximal
values $\lambda_0(\frac{1}{4})\simeq -2.2$ and
$\lambda_1(\frac{1}{4})\simeq -0.96$, and satisfy $\lambda_0(\nu)\to
-\infty$ and $\lambda_1(\nu)\to -\infty$ when $\nu\to 0$. Furthermore
we always have $\lambda_0(\nu)<\lambda_1(\nu)$. This analysis shows
that in the case of our preferred value
$\lambda=-\frac{1987}{3080}\simeq -0.64$, as well as in the cases
where $\omega_{\rm static}=-9.34$ and $\lambda=0$, the energy function
$E(\Omega)$ given by Eq. (\ref{2}), for {\it any} mass ratio $\nu$,
admits a unique extremum, which is a minimum, at some $\Omega_{\rm
ICO}$ (for corotating binaries we shall find a minimum and also a
maximum at very high frequency). We show in FIG. \ref{fig2} the graph
of $E(\Omega)$ for equal masses and $\omega_{\rm
static}=0$. Anticipating on our discussion below, it is interesting to
compare FIG. \ref{fig2} with the result of the numerical simulation
provided by the figure 16 in Ref. \cite{GGB2}.

In TABLE \ref{tab1} we present the values of the calculated frequency
$\Omega_{\rm ICO}$, the corresponding energy $E_{\rm ICO}$ and angular
momentum $J_{\rm ICO}$, at the 1PN and 2PN orders, and at the 3PN
order in the three cases where $\omega_{\rm static}=0$, $\lambda=0$,
and $\omega_{\rm static}=-9.34$. The 1PN and 2PN approximations are
defined by the obvious truncation of Eqs. (\ref{2})-(\ref{3}). Notice
how close together already are the 2PN and 3PN approximations
(however, the 1PN order seems to be quite inadequate). Let us now show
that the 3PN approximation, in standard form (Taylor approximants),
appears to be very good to locate the turning point of the ICO, in the
sense that the prediction for that point is close to the recent result
of numerical relativity.

A novel approach to the problem of the numerical computation of binary
black holes in the pre-coalescence stage, has been proposed and
implemented by Gourgoulhon, Grandcl\'ement and Bonazzola
\cite{GGB1,GGB2}. This approach uses multi-domain spectral methods
\cite{GBGM}, and is based on two approximations, the first one is
essentially ``technical'', the other one is ``physical''. The
technical assumption (which could be relaxed in future work) is the
conformal flatness of the spatial metric~: $\gamma_{ij} =
\Psi^4\delta_{ij}$. On the other hand, an imposed ``helical'' symmetry
constitutes an important physical restriction to binary systems moving
on {\it exactly} circular orbits. By helical symmetry we mean that the
space-time is endowed with a Killing vector field of the type
$\ell^\mu = \frac{\partial}{\partial t}+\Omega
\frac{\partial}{\partial \varphi}$, where $\frac{\partial}{\partial
t}$ and $\frac{\partial}{\partial \varphi}$ denote respectively the
time-like and space-like vectors that coincide asymptotically with the
coordinate vectors of an asymptotically inertial observer. A crucial
advantage of the helical symmetry, especially in view of the
comparison we want to make with the post-Newtonian calculation, is
that the orbital frequency $\Omega$ is unambiguously defined as the
rotation rate of the Killing vector. Thanks to these approximations,
Gourgoulhon, Grandcl\'ement and Bonazzola \cite{GGB1,GGB2} were able
to obtain numerically the energy and angular-momentum along the
binary's evolutionary sequence, i.e. maintaining Eq. (\ref{3'}) along
the sequence, and to determine the minimum of these functions or ICO.

The numerical calculation reported in Refs. \cite{GGB1,GGB2} has been
performed in the case of {\it corotating} black holes, which are
spinning with the orbital angular velocity $\Omega$. We must therefore
include within our post-Newtonian treatment the effect of
spins\footnote{The importance of the effect of spins in corotating
systems of neutron stars, for which the ICO is usually determined by
the hydrodynamical instability rather than by the effect of general
relativity, is well known \cite{DBSSU}.}, appropriate to two Kerr
black holes rotating at the orbital rate $\Omega$. By combining the
formula of Christodoulou and Ruffini~: $m^2=m_{\rm
irr}^2+\frac{S^2}{4m_{\rm irr}^2}$, with the known relation between
the black-hole spin and its angular velocity\footnote{More precisely
the angular velocity is defined as the one of the outgoing photons
that remain for ever at the location of the horizon; see Eq. (33.42b)
in Ref. \cite{MTW}.}~:
$S=2m^3\Omega\Big[1+\sqrt{1-\frac{S^2}{m^4}}~\!\Big]$, we obtain the
total mass $m$ and spin $S$ of each of the corotating black holes in
terms of their irreducible mass $m_{\rm irr}$,

\begin{mathletters}\label{8}\begin{eqnarray}
m &=& \frac{m_{\rm irr}}{\sqrt{1-4(m_{\rm irr}\Omega)^2}} \simeq
m_{\rm irr}+2m_{\rm irr}^3\Omega^2\;,\label{8a}\\ S &=&
\frac{4m_{\rm irr}^3\Omega}{\sqrt{1-4(m_{\rm irr}\Omega)^2}} \simeq
4m_{\rm irr}^3\Omega\label{8b}\;.
\end{eqnarray}\end{mathletters}
The irreducible masses are precisely the ones which are held constant
along the evolutionary sequences calculated numerically in
Refs. \cite{GGB1,GGB2}. Therefore our first task is to replace all the
masses parametrizing the sum $M+E$, where $M=m_1+m_2$ is the total
rest mass-energy and $E$ is the 3PN binding energy given by
Eq. (\ref{2}), by their equivalent expressions, following
Eq. (\ref{8a}), in terms of the two irreducible masses. It is clear
that the leading contribution is that of the kinetic energy of the
spins and will come from the replacement of the rest mass-energy $M$;
from Eq. (\ref{8a}) we see that this effect will be of order
$\Omega^2$ in the case of corotating binaries, which means by
comparison with Eq. (\ref{2}) that it is equivalent to an ``orbital''
effect at the 2PN order. Higher-order corrections in Eq. (\ref{8a})
will behave at least like $\Omega^4$ and correspond to the 5PN order
at least, negligible for the present purpose. In addition there will
be a subdominant contribution, of order $\Omega^{8/3}$ or 3PN, coming
from the replacement of the masses into the ``Newtonian'' part,
$\propto \Omega^{2/3}$, of the binding energy $E$ [see
Eq. (\ref{2})]. At the 3PN approximation we do not need to replace the
masses into the post-Newtonian corrections in $E$. Our second task is
to include the relativistic spin-orbit (S.O.)  interaction. In the
case of spins $S_1$ and $S_2$ aligned parallel to the orbital angular
momentum (and right-handed with respect to the sense of motion) the
S.O. contribution to the energy reads \cite{BOC,KWWspin}

\begin{equation}\label{9}
E_{\rm S.O.}= -\nu M\left(M \Omega\right)^{5/3}
 \Bigg[\left(\frac{4}{3}\frac{m_1^2}{M^2}+\nu\right)\frac{S_1}{m_1^2}
 + \left(\frac{4}{3}\frac{m_2^2}{M^2}+\nu\right)\frac{S_2}{m_2^2}\Bigg]\;.
\end{equation}
As can immediately be infered from $S\simeq 4m^3\Omega$, which is
deduced from Eq. (\ref{8b})\footnote{The moment of inertia of the Kerr
black hole in the limit of slow rotations is $I=4m^3$, in accordance
with Eq. (2.61) in Ref. \cite{paradigm}.}, in the case of corotating
black-holes the S.O. effect is of order 3PN and therefore must be
retained at the present accuracy [with this approximation, the masses
in Eq. (\ref{9}) can be chosen to be the irreducible ones]. By
contrast, the spin-spin (S.S.) interaction turns out to be much
smaller, equivalent to the 5PN order for corotating
systems. Considering all the contributions present with the 3PN
accuracy, we thus obtain three terms~: $(2-6\nu)\left(M
\Omega\right)^2$ coming from the kinetic energy of the corotating
spins; $\left(-\frac{2}{3}\nu+\nu^2\right)\left(M \Omega\right)^{8/3}$
due to a coupling between the spin kinetic energy and the orbital
energy; $\left(-\frac{16}{3}\nu+12\nu^2\right)\left(M
\Omega\right)^{8/3}$ due to the S.O. interaction
(\ref{9}). Numerically the kinetic energy of the spins will dominate
the other effects. Hence the supplementary energy that is due
specifically to the corotation reads

\begin{equation}\label{10}
\frac{E^{\rm corot}(\Omega)}{M} = (2-6\nu)\left(M
\Omega\right)^2+\left(-\frac{18}{3}\nu+13\nu^2\right)\left(M
\Omega\right)^{8/3}\;.
\end{equation}
The total binding energy of the corotating binary is the sum of
Eqs. (\ref{2}) and (\ref{10}). Notice that we must now understand all
the masses in Eqs. (\ref{2}) and (\ref{10}) as being the irreducible
masses (we no longer indicate the superscripts ``irr''), which stay
constant when the binary evolves following Eq. (\ref{3'}).

\begin{figure}[htbp]
\centerline{\epsfxsize=14cm \epsfbox{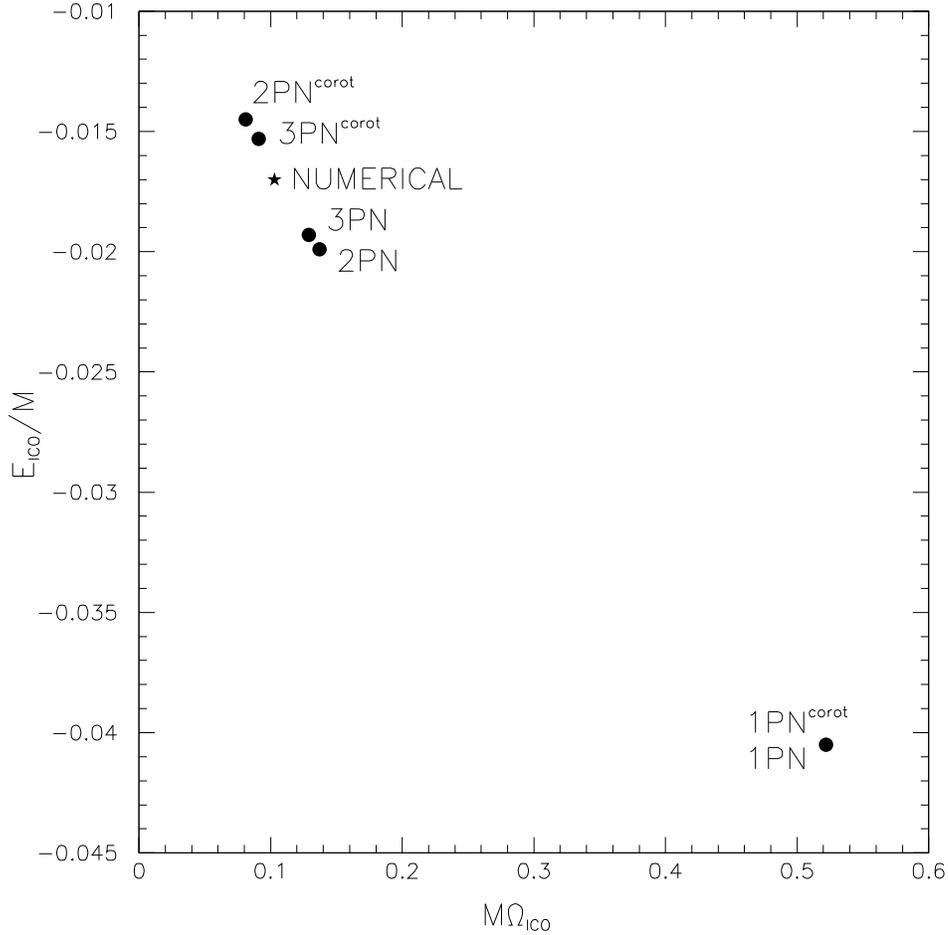}}
\vspace{0.5cm}
\caption{Results for $E_{\rm ICO}$ versus $\Omega_{\rm ICO}$ in the
equal-mass case. The asterisk marks the result calculated by numerical
relativity. The points indicated by 1PN, 2PN and 3PN are computed from
Eq. (\ref{2}), and correspond to irrotational binaries. The points
denoted by 1PN$^{\rm corot}$, 2PN$^{\rm corot}$ and 3PN$^{\rm corot}$
come from the sum of Eqs. (\ref{2}) and (\ref{10}), and describe
corotational binaries. Both 3PN are 3PN$^{\rm corot}$ are shown for
$\omega_{\rm static}=0$.}
\label{fig3}
\end{figure}

In TABLE \ref{tab2} we present our results for $E_{\rm ICO}$ and
$\Omega_{\rm ICO}$ of a corotational binary. Since $E^{\rm corot}$,
given by Eq. (\ref{10}), is at least of order 2PN, the result for
1PN$^{\rm corot}$ is the same as for 1PN in the irrotational case;
then, obviously, 2PN$^{\rm corot}$ takes into account only the leading
2PN corotation effect (i.e. the kinetic energy of the spins), while
3PN$^{\rm corot}$ involves also, in particular, the corotational
S.O. coupling at 3PN order. In FIG. \ref{fig3} we plot $E_{\rm ICO}$
versus $\Omega_{\rm ICO}$, computed with and without the corotation
effect, and compare the values with the result obtained by numerical
relativity under the assumption of helical symmetry \cite{GGB2}. As we
can see the 3PN points, and even the 2PN ones, are rather close to the
numerical value. As expected, the best agreement is for the 3PN
approximation and in the case of corotation\footnote{We have checked
that our best value, given by 3PN$^{\rm corot}$, is not significantly
modified numerically when we add the higher-order spin effects in
Eq. (\ref{10}) up to the 5PN order, i.e. including, in particular, the
S.S. interaction.}~: i.e. the point 3PN$^{\rm corot}$. However, the
1PN approximation is clearly not precise enough, but this is not very
surprising in this highly relativistic regime where the orbital
velocity reaches $v\sim (M\Omega_{\rm ICO})^{1/3}\sim
0.5$. Summarizing, we find that the location of the ICO computed by
numerical relativity, under the helical-symmetry approximation, is in
good agreement with post-Newtonian predictions. This was already
pointed out in Ref. \cite{GGB2} from the comparison with Pad\'e and
EOB methods. This constitutes an appreciable improvement of the
previous situation, because we recall that the earlier estimates of
the ICO in post-Newtonian theory~: $M\Omega_{\rm ICO}\simeq 0.06$ and
$E_{\rm ICO}/M\simeq -0.009$ \cite{KWW}, and numerical relativity~:
$M\Omega_{\rm ICO}\simeq 0.17$ and $E_{\rm ICO}/M\simeq -0.024$
\cite{Pfeiffer,Baumgarte}, strongly disagree with each other, and do
not match with the present 3PN results (see Ref. \cite{GGB2} for
further discussion).

\begin{figure}[htbp]
\centerline{\epsfxsize=14cm \epsfbox{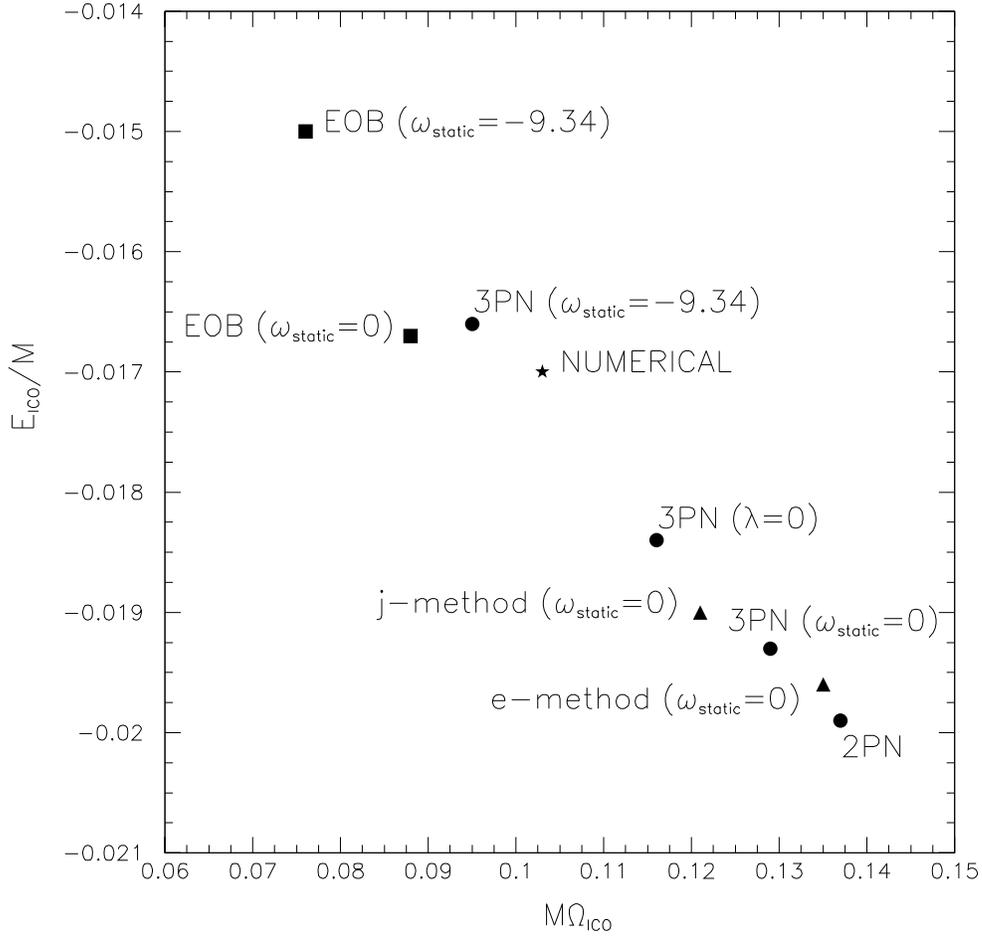}}
\vspace{0.5cm}
\caption{Results for $E_{\rm ICO}$ in terms of $\Omega_{\rm ICO}$ in
the equal-mass case. The $e$ and $j$-methods are Pad\'e approximants
at the 3PN order. EOB refers to the effective-one-body approach at the
3PN order. The points marked by 2PN and 3PN correspond to the standard
Taylor post-Newtonian series (this work). The results for Pad\'e, EOB
and Taylor are for irrotational binaries.}
\label{fig4}
\end{figure}

\begin{figure}[htbp]
\centerline{\epsfxsize=14cm \epsfbox{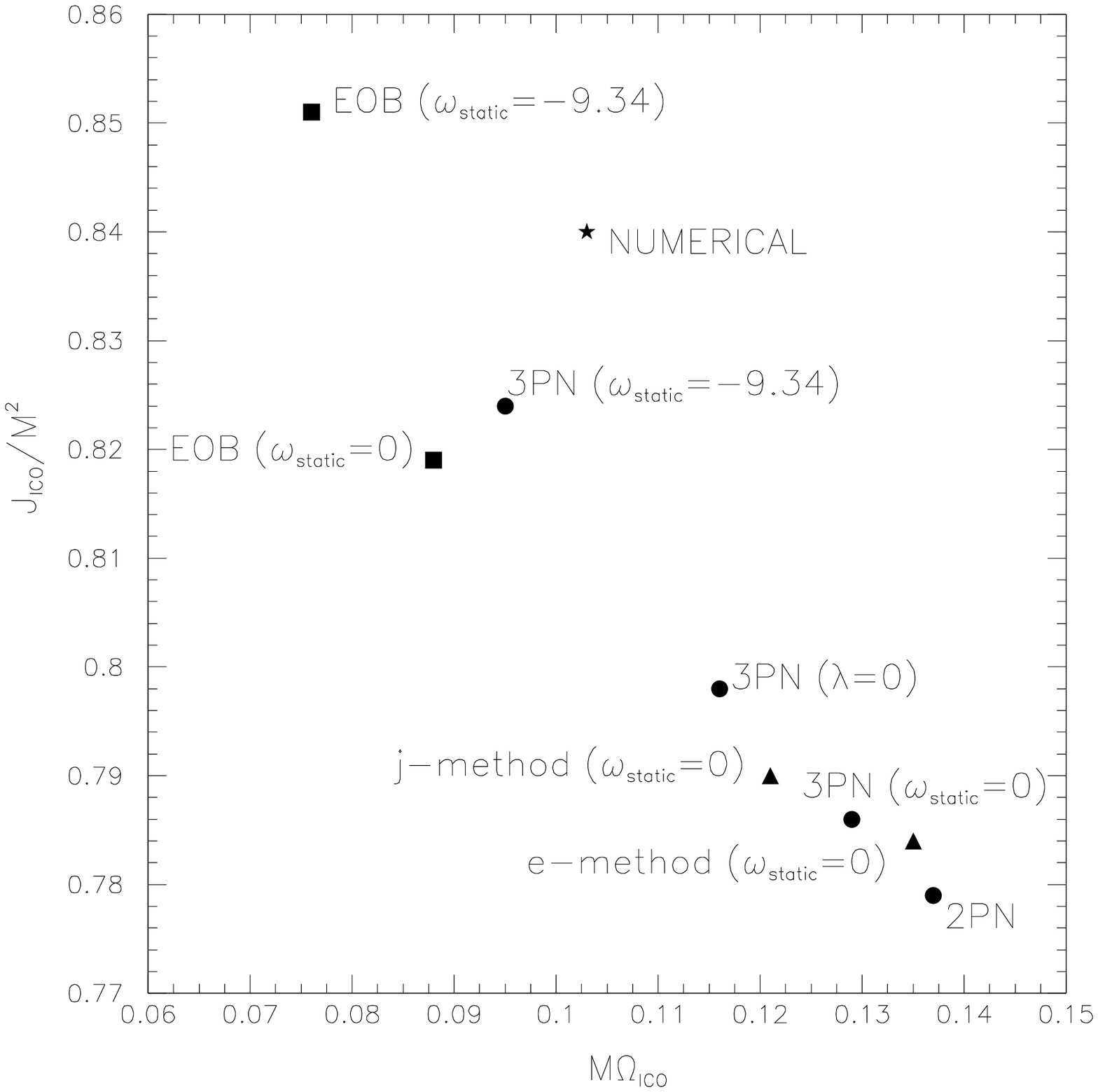}}
\vspace{0.5cm}
\caption{Same as FIG. \ref{fig4} but for the angular-momentum $J_{\rm
ICO}$.}
\label{fig5}
\end{figure}

Let us emphasize that our computation has been based on the standard
post-Newtonian approximation, expanded in the usual way as a Taylor
series in the frequency-related parameter $x = \left(M
\Omega\right)^{2/3}$ [see Eqs. (\ref{2})-(\ref{3}) and (\ref{10})],
without using any resummation techniques. In
FIGS. \ref{fig4}-\ref{fig5} we display our Taylor-series-based values
for $E_{\rm ICO}$ and $J_{\rm ICO}$ (they are indicated by the marks
2PN and 3PN), and contrast them with some results obtained by means of
resummation techniques at the 3PN order~: Pad\'e approximants
\cite{DIS98,DJSisco} and effective-one-body (EOB) methods
\cite{BD,DJSisco}. All these results agree rather well with each
other, and, as we have seen, even the 2PN (Taylor) approximation does
well.

A point we make is that the sophisticated Pad\'e approximants give
about the same results as the standard post-Newtonian expansion, based
on the much simpler Taylor approximants~: indeed, see in
FIGS. \ref{fig4}-\ref{fig5} the points referred to as the $e$ and
$j$-methods, which are 3PN Pad\'e resummations built respectively on
the energy and angular-momentum \cite{DJSisco}. For the case at hands
-- equal-mass binaries --, there is apparently no improvement from
using Pad\'e approximants. Nevertheless, it is true that in the
test-mass limit $\nu\to 0$ the Pad\'e series converges rapidly toward
the exact result \cite{DIS98}. For instance, the Pad\'e constructed in
this case from the 2PN approximation of the energy already coincides
with the exact expression for the Schwarzschild metric [given by
Eq. (\ref{3a''})]. But, the results of FIGS. \ref{fig4}-\ref{fig5}
suggest that this interesting feature of the Pad\'e approximants is
lost when we turn on $\nu$ and consider the equal-mass case
$\nu=\frac{1}{4}$. Notice also that the 2PN versions of these Pad\'e,
which are given in the table I of Ref. \cite{DJSisco}, differ much
more significantly from the corresponding 3PN ones than in the case of
Taylor. For instance, the 2PN $e$-method yields the values
$M\Omega_{\rm ICO}\simeq 0.09$ and $E_{\rm ICO}/M\simeq -0.016$, which
respectively differ by about 36\% and 22\% with the frequency and
energy given by the $e$-method at 3PN. In the case of Taylor, the same
figures are only 6\% and 3\%. Thus, on the point of view of the
``Cauchy criterium''\footnote{The Cauchy criterium for the series
$\sum a_n$ is the fact that $|a_n-a_m|\to 0$ for any $n$ and $m$.},
the Taylor series seems to converge better that the Pad\'e
approximants (for equal masses).

It is a pleasure to thank Eric Gourgoulhon for informative
discussions, Alessandra Buonanno and Gilles Esposito-Far\`ese for
useful remarks, and a referee for valuable comments.

%\pagebreak

%\narrowtext 

\begin{table}
\begin{tabular}{lcccc}
&Newtonian&1PN&2PN&3PN\\[2mm]\hline 
$\nu=\frac{1}{4}$ $\quad\omega_{\rm static}=0$&1&-0.77&-2.78&-0.97\\
$\nu=0$ &1&-0.75&-3.37&-10.55\\
\end{tabular}
\vspace{0.5cm}
\caption{Numerical values of the sequence of coefficients of the
post-Newtonian series composing the energy function (\ref{2}).}
\label{tab0}
\end{table}
\vspace{1cm}

\begin{table}
\begin{tabular}{lccc}
&$M\Omega_{\rm ICO}$&$\frac{\textstyle{E_{\rm
ICO}}}{\textstyle{M}}$&$\frac{\textstyle{J_{\rm ICO}}}
{\textstyle{M^2}}$\\[2mm]\hline 
1PN &0.522&-0.0405&0.621\\
2PN &0.137&-0.0199&0.779\\
3PN $\quad\omega_{\rm static}=0$&0.129&-0.0193&0.786\\ 
3PN $\quad\lambda=0$&0.116&-0.0184&0.798\\ 
3PN $\quad\omega_{\rm static}=-9.34$&0.095&-0.0166&0.824\\
\end{tabular}
\vspace{0.5cm}
\caption{Parameters for the ICO of equal-mass ($\nu=\frac{1}{4}$)
binary systems.}
\label{tab1}
\end{table}
\vspace{1cm}

\begin{table}
\begin{tabular}{lcc}
&$M\Omega_{\rm ICO}$&$\frac{\textstyle{E_{\rm
ICO}}}{\textstyle{M}}$\\[2mm]\hline 
1PN$^{\rm corot}$ &0.522&-0.0405\\
2PN$^{\rm corot}$ &0.081&-0.0145\\
3PN$^{\rm corot}$ $\quad\omega_{\rm static}=0$&0.091&-0.0153\\ 
\end{tabular}
\vspace{0.5cm}
\caption{Parameters for the ICO of corotational equal-mass binary
systems.}
\label{tab2}
\end{table}

\pagebreak

\pagebreak

\pagebreak

\pagebreak


\begin{references}
\bibitem{OO}T. Ohta, H. Okamura, T. Kimura and K. Hiida,
Prog. Theor. Phys. {\bf 50}, 492 (1973); {\it ibid}. {\bf 51}, 1220
(1974); {\it ibid}. {\bf 51}, 1598 (1974).
\bibitem{S85}G. Sch\"afer, Ann. Phys. (N.Y.) {\bf 161}, 81 (1985).
\bibitem{JaraS}P. Jaranowski and G. Sch\"afer, Phys. Rev. D{\bf 57},
7274 (1998); {\it ibid}. {\bf 60}, 124003 (1999); Annalen Phys. {\bf
9}, 378 (2000).
\bibitem{DJS}T. Damour, P. Jaranowski and G. Sch\"afer,
Phys. Rev. D{\bf 62}, 021501R (2000); {\it ibid}. {\bf 63}, 044021 (2001).
\bibitem{BFP98}L. Blanchet, G. Faye and B. Ponsot, Phys. Rev. D{\bf 58},
124002 (1998). 
\bibitem{BF}L. Blanchet and G. Faye, Phys. Lett. A {\bf 271},
58 (2000); Phys. Rev. D{\bf 63}, 062005 (2001).
\bibitem{BFreg}L. Blanchet and G. Faye, J. Math. Phys. {\bf 41}, 7675
(2000); {\it ibid}. {\bf 42}, 4391 (2001).
\bibitem{ABF01}V. de Andrade, L. Blanchet and G. Faye, Class. Quantum
Grav. {\bf 18}, 753 (2001).
\bibitem{DJSisco}T. Damour, P. Jaranowski and G. Sch\"afer,
Phys. Rev. D{\bf 62}, 084011 (2000).
\bibitem{DJSdim}T. Damour, P. Jaranowski and G. Sch\"afer, 
Phys. Lett. B{\bf 513}, 147 (2001).
\bibitem{DIS98} T. Damour, B.R. Iyer and B.S. Sathyaprakash,
Phys. Rev. D{\bf 57}, 885 (1998).
\bibitem{BD}A. Buonanno and T. Damour, Phys. Rev. D{\bf 59},
084006 (2001).
\bibitem{3mn}C. Cutler, T.A. Apostolatos, L. Bildsten, L.S. Finn,
E.E. Flanagan, D. Kennefick, D.M. Markovic, A. Ori, E. Poisson,
G.J. Sussman and K.S. Thorne, Phys. Rev. Lett. {\bf 70}, 2984 (1993).
\bibitem{P95}E. Poisson, Phys. Rev. D{\bf 52}, 5719 (1995); {\it
ibid}. {\bf 55}, 7980 (1997).
\bibitem{GGB1}E. Gourgoulhon, P. Grandcl\'ement and S. Bonazzola,
Phys. Rev. D{\bf 65}, 044020 (2002).
\bibitem{GGB2}P. Grandcl\'ement, E. Gourgoulhon and S. Bonazzola,
Phys. Rev. D{\bf 59}, 044021 (2002).
\bibitem{GBGM}P. Grandcl\'ement, S. Bonazzola, E. Gourgoulhon and
J.-A. Marck, J. Comput. Phys. {\bf 170}, 231 (2001).
\bibitem{DBSSU}M.D. Duez, T.W. Baumgarte, S.L. Shapiro, M. Shibata and
K. Uryu, Phys. Rev. D{\bf 65}, 024016 (2001).
\bibitem{MTW}C.W. Misner, K.S. Thorne and J.A. Wheeler, {\it
Gravitation}, Freeman (1973).
\bibitem{BOC}B.M. Barker and R.F. O'Connell, Gen. Relat. and
Gravit. {\bf 11}, 149 (1973).
\bibitem{KWWspin}L.E. Kidder, C.M. Will and A.G. Wiseman,
Phys. Rev. D{\bf 47}, R4183 (1993).
\bibitem{paradigm}K.S. Thorne, R.H. Price, D.A. Macdonald, {\it Black
holes~: the membrane paradigm}, Yale University Press (1986).
\bibitem{KWW}L.E. Kidder, C.M. Will and A.G. Wiseman, Phys. Rev. D{\bf 47},
3281 (1993).
\bibitem{Pfeiffer}H.P. Pfeiffer, S.A. Teukolsky and G.B. Cook, 
Phys. Rev. D{\bf 62}, 104018 (2000).
\bibitem{Baumgarte}T.W. Baumgarte, Phys. Rev. D{\bf 62}, 024018 (2000).
\end{references}
\end{document}